# Generative AI for Object-Oriented Programming: Writing the Right Code and Reasoning the Right Logic

Gang Xu[1], Airong Wang[2], Yushan Pan[1]

*Abstract*— We find ourselves in the midst of an explosion in artificial intelligence research, particularly with large language models (LLMs). These models have diverse applications spanning finance, commonsense knowledge graphs, medicine, and visual analysis. In the world of Object-Oriented Programming (OOP), a robust body of knowledge and methods has been developed for managing complex tasks through object-oriented thinking. However, the intersection of LLMs with OOP remains an underexplored territory. Empirically, we currently possess limited understanding of how LLMs can enhance the effectiveness of OOP learning and code writing, as well as how we can evaluate such AI-powered tools. Our work aims to address this gap by presenting a vision from the perspectives of key stakeholders involved in an OOP task: programmers, mariners, and experienced programmers. We identify critical junctures within typical coding workflows where the integration of LLMs can offer significant benefits. Furthermore, we propose ways to augment existing logical reasoning and code writing, ultimately enhancing the programming experience.

## I. INTRODUCTION

The publication of the IEEE Spectrum, "How Coders Can Survive- and Thrive- in a ChatGPT World," by Rina Diane Caballar[3] in 2023 marked important tips for programmers to stay ahead of generative AI, advising that artificial intelligence can be used augment programming. GitHub Copilot[7], ChatGPT[8], and Google Bard[17] can act as AI-powered coding assistants to augment programming and provide suggestions as programmer codes. In particular, programmers can also use ChatGPT and Google's Bard to call application programming interfaces or generate code snippets[3].

The advancements in Large Language Models (LLMs) serve as the foundation for Generative AI (GenAI) models and the recent surge in creative support tools[16]. These developments offer promising opportunities to enhance real-world professional programming practices[21]. Online coaching platforms like W3Schools are now utilizing ChatGPT 4 to provide students with a learning experience akin to having an experienced programmer's assistance[25]. Additionally, platforms like AlphaCode[11] have surpassed earlier AI code writing standards, including Codex, released by OpenAI in 2021. While exciting, it is unknown at this point if these tools are improving human performance, and if so, in what ways[15]. What interfaces and interactions will best support professionals remains an open question[1][11].

*This work was supported by XJTLU TDF22/23-R26-219.
[1]G. XU and Y. Pan is with Department of Computing, Xi'an Jiaotong-Liverpool University, 215123 Suzhou, China yushan.pan@xjtlu.edu.cn
[2]A. Wang is with School of Language, Xi'an Jiaotong-Liverpool University, 215123 Suzhou, China airong.wang@xjtlu.edu.cn

Designing a software product is a labor-intensive task. While numerous AI-based coding assistants can aid in code completion and generation, the fundamentals of programming remain essential. Programmers must still possess the ability to read and reason about their own and others' code, understanding how their code fits into larger systems. Moreover, one of the most crucial programming skills is domain knowledge, including domain problem-solving. Analyzing domain problems and devising elegant solutions remain highly valued coding expertise. Therefore, it's important to discern the programmer's role and the advantages of AI. It's vital to remember that Generative AI (GenAI) primarily generates statistical outputs from a large model. There are distinctions between what a programmer does and what the model produces; being a human programmer involves more than just writing code[14]. Software engineering practices offer additional value, encompassing system design and software architecture. Human coders are responsible for determining code structure, suitable abstractions, and interface requirements. The challenge is how to leverage GenAI to produce correct code and logical reasoning, collaborating with human programmers to create fully functional systems[2].

In this context, we contribute to the discussion on human-centric computing challenges when applying GenAI to achieve both correct code writing and accurate, logical reasoning in code. Regarding code writing, we identify a potential tradeoff between inspiration and efficiency in generating high-quality code. Additionally, we discuss the challenges posed by GenAI tools in specific domains, such as the maritime industry. In terms of logic reasoning, we address both human-centered challenges in collaborative work with GenAI-based logical reasoning and technical challenges involving consumer preference's influence on the software development process.

The rest of the paper is structured as follows: Section 2 covers related work, Section 3 presents a case involving less-experienced programmers designing a collaborative tool for mariner tasks called 'Timing.' Section 4 outlines our methodology, followed by our findings in Section 5. Section 6 provides a discussion, and the paper concludes in Section 7.

## II. RELATED WORK

### A. GenAI in Software Engineering

With the growing popularity of GenAI, it appears that almost anyone can become a programmer. People believe that all it takes is talking to a computer, and it will generate the necessary code based on your descriptions. Admittedly, generative AI's ability to comprehend natural language and translate it into code is a powerful tool, enabling developers to describe their code requirements, with the AI then generating the corresponding code. However, it's essential to understand that current GenAI is not yet capable of generating an entire code repository for a complete software product. Therefore, the consensus is that generative AI enhances developer productivity and assists in focusing on critical development aspects rather than replacing programmers altogether. Essentially, people see generative AI as a developer productivity booster[19].

Using AI to support development follows a path of specifying systems at a more abstract level, such as using writing prompts to describe systems, which has the potential to accelerate development and broaden the developer base. What remains uncertain is the quality of the results, the legal implications of this approach, and its impact on developer skills. It's also vital to demystify the notion that GenAI's ability to generate code is magical. The mechanism behind it is relatively straightforward. Knowledge resides in books, documents, and human minds until it is intentionally applied[6]. The act of intentionality is where challenges often arise; sometimes, we might not recall or locate the right piece of knowledge to address a specific problem. Generative AI can address this issue by efficiently ingesting data, learning, and generating knowledge. This mechanism allows us to access our knowledge capital exactly when and where we need it, in a practical and timely manner.

Notably, AI can also enhance software testing, which relies on a deep understanding of the product under test, technical aspects, and customer use cases. Effective testers are those who possess a thorough understanding of how the product works under the hood and how it can be used. When we attempt to automate test cases, issues related to script rigidity can arise; a minor change in a resource ID can lead to the failure of an entire test case[20].

In software development, programming, knowledge, and testing represent the three key pillars that have the potential to transform how software engineers work: quickly generating ideas and solutions, generating code snippets, and generating tests while maintaining high productivity and effectiveness. Companies should seriously consider integrating generative AI into their strategies and roadmaps. With generative AI as part of a company's strategy, software engineers can become more efficient problem solvers and produce high-quality products rapidly[5].

While research into machine learning algorithms continues to yield astonishing possibilities, the widespread use of these algorithms shifts the focus toward integration, maintenance, and the evolution of AI-driven systems. Despite the variety of machine learning frameworks available, there is limited support for process automation and DevOps in machine learning-driven projects. Some studies suggest that metamodels can support the development of deep learning frameworks, dealing with the increasing variety of learning algorithms. Authors have proposed a deep-learning-oriented artifact model that serves as a foundation for automation and data management in iterative, machine learning-driven development processes, including schema and reference models to structure and maintain versatile deep learning frameworks[13].

*B. GenAI in Interaction Studies*

In addition to software engineering, numerous studies focus on generating tests and images using AI/GenAI systems[10][26]. Image generation has gained popularity with machine learning models creating images from text descriptions. For instance, researchers have explored applications in painting[18][23] and creative arts[22].

Teo Sanchez[10] has investigated how users generate images from text, contributing to the understanding that GenAI serves as a recreational activity, predominantly engaging narrow sociodemographic groups who utilize auxiliary techniques across various platforms and beyond simple request-response interactions.

Generative AI can also play a crucial role in the creative design process, particularly in ideation, early prototyping, and sketching. For example, Tholander and Jonsson conducted a workshop to develop design concepts using GPT-3. Their findings indicate that GPT-3 can enhance three aspects of user experiences: practical usefulness and limitations in design ideation processes, the influence of user interaction on shaping expectations regarding the system's capabilities and potentials, and co-creative processes involving both humans and AI in post-human perspectives on design and technology use[24].

Similar studies can be found in other interaction design communities, such as generative AI systems for visual storytelling aimed at young learners[9]. Previous research has shown the significant utility of using text prompts and how AI-augmented workflows can boost productivity in creative tasks for end users. However, the definition of multimodal interactions beyond text prompts is an area that needs further exploration to enable end users to have rich points of interaction for truly co-piloting AI-generated content creation [12].

*C. Recap the Related Work*

GenAI doesn't merely categorize data or interpret text using predefined models; instead, it has the capability to generate entirely new content, ranging from code and images to molecules and designs. This shift in AI's functionality moves us from a focus on problem-solving to problem-finding, essentially changing the AI's 'role' from decision-maker to human support.

However, several key questions remain unanswered. It's unclear how humans perceive and make sense of generative AI algorithms and their outputs. Furthermore, the issue of controlling and interacting with these powerful capabilities remains unresolved. The nature of collaboration between creative humans and creative technologies is also uncertain, and we're yet to understand the patterns that will emerge in such partnerships.

Given these uncertainties, it's crucial to explore the intersection of the interdisciplinary research domain of generative AI and Human-Computer Interaction (HCI). This exploration can shed light on opportunities to enhance the field by addressing aspects like accountability, transparency, and fairness.

## III. THE CASE

This section addresses safety concerns within a maritime workplace where a group of maritime operators engages in dynamic positioning, automation information systems, calculations, and alarm clock usage. Dynamic positioning systems are crucial components in embedded systems on various types of ships, particularly special-purpose vessels. They are designed to ensure safety through precise positioning. Operators face challenges due to adverse working conditions such as waves, winds, and the risk of collisions with structures like offshore oil platforms. Improper operation of dynamic positioning systems can disrupt the ship's balance1.

The current system design requires operators to manually collaborate through observation and timekeeping. This manual approach helps the ship maintain balance during interactive operations with offshore platforms and liquid transfers on the ship's deck.

To tackle this issue, we've enlisted the expertise of four software developers, one of whom has extensive experience in the IT industry, while the other three are senior PhD students with IT industry experience. These engineers are collaborating with researchers to create collaborative systems supporting local maritime companies. With the rise in popularity of ChatGPT in 2023, they aim to enhance their work and reduce their workload. Leveraging AI tools for programming tasks is a logical choice in their current context.

Although they have experience with GitHub Copilot, it primarily focuses on code completion and lacks the capacity to answer a broader range of questions outside typical programming workflows. Furthermore, these software developers have no prior experience in the maritime domain. Therefore, they have opted for ChatGPT to gain knowledge about this domain and seek answers to questions that might influence their code frames and snippets. ChatGPT is the preferred choice for this project.

## IV. METHOD

Central to our research is the development of functional prototypes to better understand interactions and collaboration systems by utilizing ChatGPT 4.0, with a specific focus on collaborative writing with AI. This approach enables us to explore realistic interactions. That said, our primary methods for understanding software developers' motivations and perspectives on current GenAI tools, including their advantages and limitations, are through observations and interviews.

To minimize the potential influence of one software engineer's opinions on others, we conducted individual interview sessions. This approach was chosen because the software engineer who is still working in a large IT industry may have unique perspectives from the industry, which might not be widely known among software engineers who have been primarily engaged in research in recent years.

Each interview lasted between 1 hour and 1.5 hours, with an average duration of approximately one hour. All interviews were conducted between May 2023 and October 2023. As a token of our appreciation for their time and dedication, each software engineer received a Starbucks gift voucher valued at 200 Chinese Yuan. Due to geographical constraints, observations were conducted individually. The first author actively participated in observations, taking notes during the process. All audio interviews and video recordings were conducted with the software engineers' consent and approval. Our study received ethical approval from the institutional ethical committee, in accordance with the university's regulations.

All interview transcripts and observation notes were extensively discussed within the research team, leading to a consensus. We performed a thematic analysis of the transcripts, creating codes for data analysis. From each software developer's perspectives, we categorized themes, identified overarching themes, and extracted design implications, which we will explain in the following section.

## V. FINDINGS

In this section, we will present the findings of our current study, which can be categorized into three primary areas: team governance in development, the code design process, and copyright issues.

### A. Team Governance on Systems Development

To gain an overview of the development process, team members must adhere to a standard procedure. For instance, the four software engineers opted for agile development to adapt and customize collaborative tasks at sea, while also incorporating the ChatGPT prompt engineering principles (as discussed in [4]). In the design of collaborative systems, software engineers must prioritize fairness, human-centered values, and the well-being of HSE (Health, Safety, and Environment) to ensure that the final product caters to the diverse expertise and backgrounds of maritime operators throughout the software development lifecycle.

Additionally, ethics should be a fundamental component within the development team to promote a more ethical approach to collaborative system development. These concerns have been thoroughly discussed among the software engineers and the researchers involved in the requirements engineering process.

*1) Requirements Analysis:* To conduct requirements analysis, software engineers hold meetings exclusively via Zoom. During these sessions, they collectively review the case and engage in discussions regarding specific scenarios within the development process. One of the engineers is tasked with utilizing prompt engineering to interact with ChatGPT, while the remainder of the team participates in discussions.

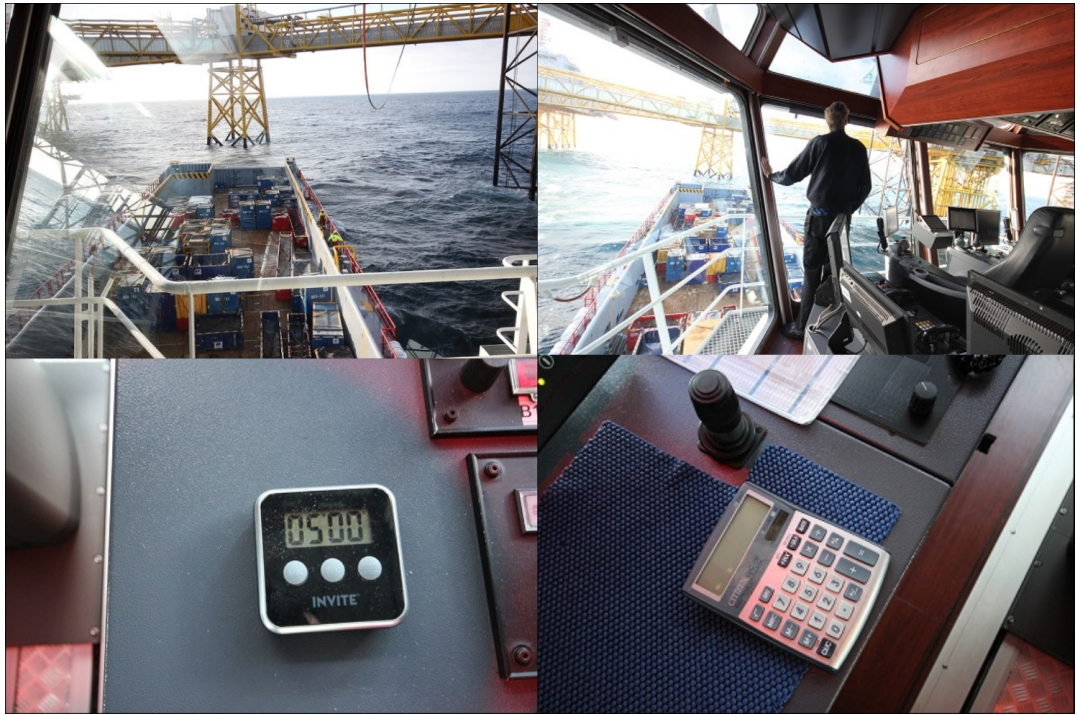

Fig. 1. The collaboroative work at sea

It was observed that this engineer is well-acquainted with the ChatGPT interface and skillfully carries out interactions. For instance, when discussing collaborative work, the engineer inputs a paragraph that provides a comprehensive overview of the tasks typically undertaken by maritime operators in their daily work at sea. The engineer further explains:

> I typed in the texts because we believe that, as we engage in conversation, the texts we input provide ChatGPT with an imaginary 'role' to better comprehend the project. For example, we might say, 'Now, my master asks me to understand the work context.' This approach helps ChatGPT contextualize the project and its requirements.

Following this, the engineer enters the task: 'I want you to develop software for managing collaborative tasks.' They add, 'I also want it to be a web-based application written in Java.'

However, the response generated by ChatGPT includes a general code with some context and instructions. For instance, it says, 'Sure! To build a web application hosted as an AWS Lambda function, we will need to use API Gateway to expose our Lambda function via HTTP endpoints. I can guide you on creating a simple record collection manager using Java, AWS Lambda, and API Gateway.' However, some of the engineers find the output from ChatGPT less helpful. One engineer argues:

> Well, this isn't surprising. I told you, Lin, that you won't receive actual code; instead, you'll get a guide. It's helpful to us, but it might not be as helpful as we'd like.

In fact, other engineers also concurred with the statement made by that engineer. The output is, more or less, a structure or guidance on how to develop, rather than providing detailed information in line with the specific project analysis. As a software engineer, following ChatGPT's instructions is not always necessary. Instead, engineers still need to adhere to their own analysis of the requirements to ensure their design is comprehensive and well-considered.

*2) Requirements Specification:* ChatGPT-promoted programming is a complex endeavor. It's worth noting that GenAI systems aim to autonomously solve problems with a level of independence and agency, which cannot always be fully specified. Therefore, it is crucial to carefully define ethical requirements in a measurable way and avoid vague and unverifiable requirements.

For instance, while designing timing functions for when the ship is engaged in interactive operations with an oil platform, one engineer raised a question: What criteria should be used for liquid shifting and time measurement? Another engineer argued that maritime operators should possess basic calculation skills for this purpose. However, according to current international maritime law, ship owners have no obligation to ensure mariners possess such skills, as it falls outside the scope of the standard mariner training curriculum. Unlike marine engineering subjects typically taught at the university level, training for mariners is usually conducted in vocational schools.

This presents a challenge for software engineers in determining how to impart the necessary computing skills and who should be responsible for providing this training. Should they simply provide a prototype to the mariners to use? When this question was posed to ChatGPT, it unsurprisingly confirmed the need for mariners to have a strong understanding of liquid calculations. As one of the engineers pointed out:

> I'm unsure whom to trust. Our requirements state the need to support this type of work, but we're uncertain if the information is accurate. Whom should we consult for guidance?

### B. Construction with Reuse

The code design process is perhaps the most straightforward phase for software engineers working with ChatGPT. Software engineers craft their instructions in a conversational style, using carefully chosen active voice and incorporating rhetorical questions that prompt ChatGPT to generate relevant code. However, software engineers have raised two crucial issues. They express concerns that the generated codes may not be entirely reliable. For example, clone-and-own is a reuse practice that involves leveraging an existing product with similarities to adapt it to new requirements by adding or removing features. This ad-hoc approach to reuse is accepted in industrial settings due to its simplicity, availability, and independence from assigned developers. However, in the long term, this practice of creating multiple variants may not be suitable, as the proliferation of variants and features can significantly complicate the maintenance and management of the entire product family. In the present study, we also observed that software engineers built three Java variants. As one of the engineers inputs:

> (Software Engineer) I have three variants of a Java implementation of a collaborative systems. The first one is Timing. The second one is Calculator. And the third variant is Liquid systems. Please considering the three variants, can you summarize what is common and what varies?

The ChatGPT can provide a quites precise answer; however there is a questionable point that there is no apparent link between Consortium class and Converter class. Actually, these two classes are essential for our collaborating systems when doing timing, calculating and liquid shifting at sea for balancing the ship. Then software engineers move to the next request a template engine that would allow deriving the three original variants.

CollaborativeSystem.java Template
```
public class CollaborativeSystem {
    public void commonFunctionality() {
        // Common code that applies to all variants
    }
    // Variant-specific elements
    <#-- Variant-specific fields, methods,
    or code will be inserted here -->
}
```

Java class
```
import freemarker.template.Configuration;
import freemarker.template.Template;
import freemarker.template.TemplateException;

import java.io.File;
import java.io.FileWriter;
import java.io.IOException;
import java.util.HashMap;
import java.util.Map;

public class CodeGenerator {
    public static void main(String[] args) {
        Configuration cfg = new Configuration(Configuration.VERSION_2_3_31);
        cfg.setClassForTemplateLoading(CodeGenerator.class, "/templates");

        // Define variant-specific data
        Map<String, String> variantData = new HashMap<>();
        variantData.put("timing", "<#-- Timing-specific elements -->");
        variantData.put("calculator", "<#-- Calculator-specific elements -->");
        variantData.put("liquid_system", "<#-- Liquid system-specific elements -->");

        for (Map.Entry<String, String> entry : variantData.entrySet()) {
            String variantName = entry.getKey();
            String variantElements = entry.getValue();

            try {
                // Load the FreeMarker template
                Template template = cfg.getTemplate("collaborative_template.ftl");

                // Create the output file
                File output = new File(variantName + "_variant.java");
                FileWriter writer = new FileWriter(output);

                // Populate the template and write to the output file
                Map<String, Object> root = new HashMap<>();
                root.put("variant_elements", variantElements);
                template.process(root, writer);

                // Close the output file
                writer.close();
                System.out.println("Generated " + output.getName());
            } catch (IOException | TemplateException e) {
                e.printStackTrace();
            }
        }
    }
}
```

### C. Copyright of Coding Logic

We have observed that software engineers are particularly mindful of the language they use when communicating with ChatGPT. For instance, there was an engineer with an industry background who chose not to use ChatGPT personally. Instead, during the study, he exclusively engaged with other engineers and encouraged those who wanted to use ChatGPT to structure their questions carefully. This is an intriguing phenomenon, and when we inquired about this matter, the engineer provided the following response:

> You may be aware that most AI-based assistant tools, if not all, raise concerns about potential copyright implications for software developers. Take GitHub Copilot as an example; it utilizes the data stored on GitHub to train its AI model, and in return, provides us with Copilot to assist with coding. The more you contribute to such AI tools, the more your programming knowledge and expertise may be utilized without compensation.

In light of the engineer's comments, it's hard to argue that ChatGPT wouldn't behave similarly. ChatGPT, too, disregards statistical patterns within existing code structures, deriving these patterns through intricate probabilistic analysis of its training data. When responding to a user's request, ChatGPT generates code to achieve the intended functionality. This process makes it quite likely that the copyrights associated with the coding logic could become unpredictable, potentially leading to violations of engineers' rights. We can predict that this potential violation will not be limited to coding but could extend to various forms of creative work, including art, UI design, and many other inspirational endeavors.

## VI. DISCUSSION

In CSCW research, in addition to studying user collaboration, it is essential to reflect on the lessons learned from the research project, examine the study's reflexivity, and involve participants to validate research outcomes. Our work is no exception. In this section, we address three key insights we've gained, with the hope of facilitating engagement with other researchers who share similar interests in our current work.

*A. GenAI, Technology Strategy and Management*

ChatGPT can indeed assist programmers to some extent in their software development work. However, this model operates on a certain level of appropriating others' intellectual property and labor. While open-source software and knowledge repositories on platforms like GitHub encourage knowledge sharing and innovation, they do not imply a lack of protection. Open source promotes knowledge sharing and innovation, but it should not be abused to circumvent intellectual property laws. This lack of respect for original creators and disregard for relevant legislation is concerning.

In our research, software engineers did not generate code with specific protective significance. However, it remains unclear whether ChatGPT, in the process of forming relevant code frameworks, used models pretrained on open-source databases. This has led to the current state of code generation being unpredictable and lacking in transparency. Even though software engineers can, to some extent, adjust their interaction patterns and dialogue paradigms with ChatGPT, it's insufficient to ensure model interpretability.

Two primary driving factors are contributing to the increasing use of large AI models by software engineers. First, as market choices narrow, the variety of large AI models tends to align with the most widely used categories. Second, only a few enterprises will continue to develop their AI models. In addition to existing intellectual property protection, we must also consider privacy regulations. Our interaction data should not be exposed as certain large models phase out. This signifies the importance of distinguishing between data we own and data that requires privacy protection.

*B. Inability to Growth?*

"The best way to predict the future is to invent it." says Dr Alan Key. While the use of AI tools is criticized for potentially diminishing motivation and the spirit of innovation to some extent, through our study, we have found that software engineers still exhibit a considerable level of work motivation and interest when presented with new tools. As one young engineer expressed, his focus is not on AI itself, but rather on how AI can swiftly provide him with work-related inspiration, much like those generated code frameworks. This, to some extent, allows him to validate his own ideas and consequently boosts his self-assessment of abilities.

If a feedback mechanism is established where we anticipate that AI can work in alignment with our work paradigm, the positive feedback generated from these paradigm-driven interactions can inspire us to view our work more positively. Learning code and everyday code-writing have become feedback-based work mechanisms, and software engineers play a role within this feedback mechanism. This opens up new perspectives and methods for self-assessment and correction, helping to prevent fatigue during the code-writing and learning process.

## VII. CONCLUSION

In this paper, we empirically studiedthe intersection of Large Language Models (LLMs) with Object-Oriented Programming (OOP) represents a largely uncharted domain. Our research contributes to fill the void of code writing and logic reasoning by offering a visionary perspective from the standpoint of key stakeholders involved in OOP tasks, including programmers, mariners, and seasoned developers. Our paper proposed critical junctures within typical coding workflows where the integration of LLMs can bring substantial advantages. Furthermore, we have put our reflections to bolster existing logical reasoning and code writing, ultimately elevating the programming experience.